# Localised Thermal Emission from Topological Interfaces

M. Said Ergoktas[1,2], Ali Kecebas[3], Konstantinos Despotelis[1,2], Sina Soleymani[3],
Gokhan Bakan[1,2], Askin Kocabas[4], Alessandro Principi[5],
Stefan Rotter[6], Sahin K. Ozdemir[7], Coskun Kocabas[1,2,8*]

1. Department of Materials, University of Manchester, Manchester, M13 9PL, UK
2. National Graphene Institute, University of Manchester, Manchester, M13 9PL, UK
3. Department of Engineering Science and Mechanics, The Pennsylvania State University, University Park, Pennsylvania 16802, USA
4. Department of Physics, Koc University, Istanbul, Turkey
5. Department of Physics, University of Manchester, Manchester, M13 9PL, UK
6. Institute for Theoretical Physics, Vienna University of Technology (TU Wien), 1040 Vienna, Austria.
7. Materials Research Institute, The Pennsylvania State University, University Park, Pennsylvania 16802, USA
8. Henry Royce Institute for Advanced Materials, University of Manchester, Manchester, M13 9PL, UK

*Corresponding author. E-mail: coskun.kocabas@manchester.ac.uk

**Abstract:** The control of thermal radiation by shaping its spatial and spectral emission characteristics plays a key role in many areas of science and engineering. Conventional approaches to tailor thermal emission using metamaterials are severely hampered both by the limited spatial resolution of the required sub-wavelength material structures and by the materials' strong absorption in the infrared. Here, we demonstrate a promising new approach based on the concept of topology. By changing a single parameter of a multilayer coating, we control the reflection topology of a surface, with the critical point of zero reflection being topologically protected. As a result, the boundaries between sub-critical and super-critical spatial domains host topological interface states with near-unity thermal emissivity. Our experimental demonstration of this effect shows that topological concepts enable unconventional manipulation of thermal light with promising applications for thermal management, energy harvesting and thermal camouflage.

A very fruitful concept that has recently emerged in the field of non-Hermitian photonics(*1–3*) is that of "Coherent perfect absorption"(*4–8*), which has been conceptualized as the time-reversed process of a laser operating at its first lasing threshold(*9, 10*). In the same way as a laser cavity emits coherent light with a sufficient amount of gain, the same cavity will turn into a "Coherent Perfect Absorber (CPA)" with the same amount of negative gain (i.e., loss). Here, we extend this concept for thermal emitters, for which Kirchhoff's law of thermal radiation directly links the thermally induced emissivity (ε, gain) of a material with its absorptivity (α, loss) at thermal equilibrium(*11–13*). For broadband and incoherent radiation, this reciprocity relation has the well-known consequence that a perfect thermal emitter is a "black body" that perfectly absorbs light at all frequencies(*14*). In turn, designing a perfect thermal emitter for coherent radiation at a single frequency involves the creation of a CPA for light incident right at the desired frequency value. Moreover, for single-sided illumination, such a CPA can be realized by an optical cavity at the critical coupling condition(*8, 15*). To guarantee the coherence of the perfect emission (*ε=1*) of this critically coupled cavity, the coherence length of the thermal radiation must be longer than the length associated with a cavity round trip(*11*).

To implement the concept of a topologically protected perfect thermal emitter at a well-defined frequency, we use a lossy optical cavity as shown in Fig. 1A. In this cavity, a thin (t<20nm) platinum (Pt) layer acts both as a broad-band thermal emitter with controllable thermal emissivity and as a cavity mirror with tuneable reflectivity. We emphasize that the emissivity ε of a free-standing thin film is fundamentally restricted to values below 0.5 (see Fig. S1). However, placing the thin film near a strongly reflecting (metallic) substrate enables coherent interference of the thermal radiation emanating from the top and bottom sides of the film's surface, resulting in emissivity values larger than 0.5 at certain wavelengths. For the realization of the strongly reflecting substrate (back-mirror), we use a thick layer of gold behind a 2 µm dielectric spacing layer of Parylene-C that is transparent in the infrared. The emissivity of the back mirror (<0.05) and that of the dielectric layer have negligible effects on the overall emissivity of the cavity.

Precise tuning of the thickness of Pt film reveals unity emissivity at the critical thickness of $t_c \approx 2.3$ nm, as detailed in Fig. 1B and 1C. As we will outline here below, this condition of perfect emission is not just a coincidental phenomenon, but rather of topological origin (*16–18*). The topological aspects of the phase transition occurring when crossing the point of critical coupling becomes evident in the complex representation of the frequency-dependent Fresnel reflection coefficient $r(\omega)$, as depicted in Fig. 1D. Here, the reflection spectrum exhibits periodic resonances, represented as recurring circles in the complex plane. With extremely thin films (*t* <1nm, sub-critical coupling), the reflection spectrum forms a small circle near $r = -1$, corresponding to the reflective properties of a perfect electric conductor (PEC). At the critical thickness, $t_c \approx 2.3$ nm, the reflection spectrum intersects the origin, resulting in zero reflection and, correspondingly, unity emissivity and unity absorption. For thicknesses exceeding $t_c$ (super-critical coupling), the reflection spectrum encircles the origin, indicative of a transition to a nontrivial topological phase. As the thickness of the Pt is tuned, the surface's reflection is transitioning from resembling an electrical conductor to mimicking a magnetic conductor with non-trivial reflection topology. In this context, the topological invariant is the winding number(*19*), defined as $w = \frac{1}{2\pi}\oint d\varphi$, where the phase $\varphi = \arg[r(\omega)]$ of the Fresnel reflection coefficient is integrated here over the cavity's free spectral range (1325 cm$^{-1}$). While this phase is not defined at the origin, $r(\omega)=0$, the fact that this critical point is encircled for super-critical cavities, but not for sub-critical ones, divides the system into two different topological classes. To investigate the characteristics of thermal emission, we utilized a Fourier Transform Infrared Spectrometer (FTIR), employing the sample itself as a heat source, and measured the first-order temporal correlations of the emitted infrared light. Strikingly, we observed a significant enhancement in the visibility of the interferograms precisely at the transition point (i.e., critical coupling point), a compelling indicator of an extended coherence length of the emission (see Fig. S2).

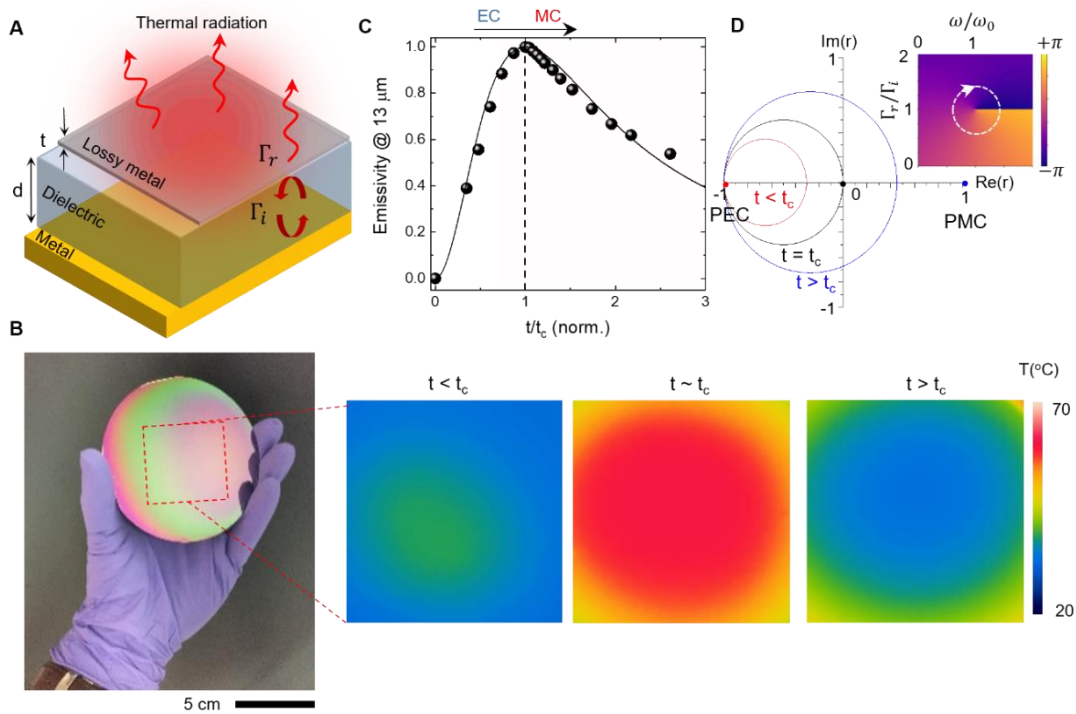

**Fig. 1. Topological phase transition in a thermal emitter**: **A,** Schematic drawing of the thermal emitting surface consisting of a lossy metal (top surface, thickness *t*) separated by a dielectric layer from a metallic substrate (bottom surface). The thermal radiation from this lossy cavity shows topologically distinct behaviour defined by the balance between radiative $\Gamma_r$ and intrinsic $\Gamma_i$ losses as controlled by the thickness *t* of the top metal film. **B,** Photograph of the sample and its infrared camera images showing the variation of thermal radiation from the surface for three different thickness. IR images recorded from the sample placed on a hot plate at 70°C. At a critical thickness ($t_c \approx 2.3$ nm) we observe near perfect thermal emissivity ($\varepsilon \approx 1$). **C,** Variation of the far-IR thermal emissivity ($\lambda = 13$ µm) as a function the thickness *t* of the lossy metal on top. The solid line shows the emissivity calculated by the model. **D,** The complex reflectivity spectra $r(\omega)$ for the three cases of sub-critical (red), critical (black) and super-critical coupling (blue). The inset shows the reflection phase in parameter space around the critical coupling point, $r_c = 0$, where a phase vortex appears, and intrinsic cavity losses are matched by the radiative losses.

Further insight is provided in the inset of Fig. 1D, which illustrates the phase singularity in the parameter space occurring when the radiative loss, $\Gamma_r$, precisely matches the cavity's intrinsic loss, $\Gamma_i$. The reflection coefficient of the cavity is calculated using the coupled mode theory(16) as $r = -1 + \frac{2\Gamma_r}{-i(\omega-\omega_0)+\Gamma_i+\Gamma_r}$ where $\omega_0$ is the resonance frequency defined by the length of the cavity. The thickness of the thermal emitting layer defines both intrinsic ($\Gamma_i \propto t$) and radiative loss ($\Gamma_r \propto \frac{1}{t}$) of the cavity. Fine tuning of the metal thickness thus enables a continuous transition from trivial to nontrivial topology in the reflection coefficient. To analyse the reflection phase, we built a phase-sensitive FTIR, to enable the direct placement of the sample within the interferometer (as depicted in Fig. 2A) and measured the reflection phase for various Pt thickness. This setup facilitates the acquisition of an interferogram, which effectively captures the first-order correlation between the reference beam and the beam reflected by the sample, thereby yielding a direct measurement of the complex Fresnel reflection coefficient. Figs. 2B and 2C represent the magnitude $|r(\omega)|$ and the phase $\varphi = \arg[r(\omega)]$ of the reflection coefficient for the infrared spectrum covering the first 5 cavity modes. Notably, the topological difference is manifested in the phase spectra; the reflection phase from the trivial surface reverts to its original state, while the non-trivial surface exhibits a cumulative phase increment of $2\pi$ for each resonance. This $2\pi$-phase is an analogue of the Berry phase in Su–Schrieffer–Heeger model which is a one-dimensional lattice model with topological features(20). The critical thickness shows slight variation for different modes due to the wavelength dependence of intrinsic loss of the cavity. Further, in Fig. 2D, we present the complex reflection spectrum for the third mode, comparing the results for the two distinct topologies. Fig. 2E shows the singularity of the zero reflection in the parameter space. The phase map, displaying φ versus ω with periodic boundary conditions for both axes, is homeomorphic to the surface of a torus with genus one as depicted in Fig. 2F. When the phase spectrum of reflection is represented on a toroidal surface, it offers a distinct method to visualize topological differences. The red and blue curves represent the phase spectra of trivial and nontrivial surfaces, respectively. These curves, which are not isotopic on the torus surface ($\mathbb{R}^2$), illustrates the differing topologies. They cannot be continuously mapped onto each other (see Fig. S3 for more details).

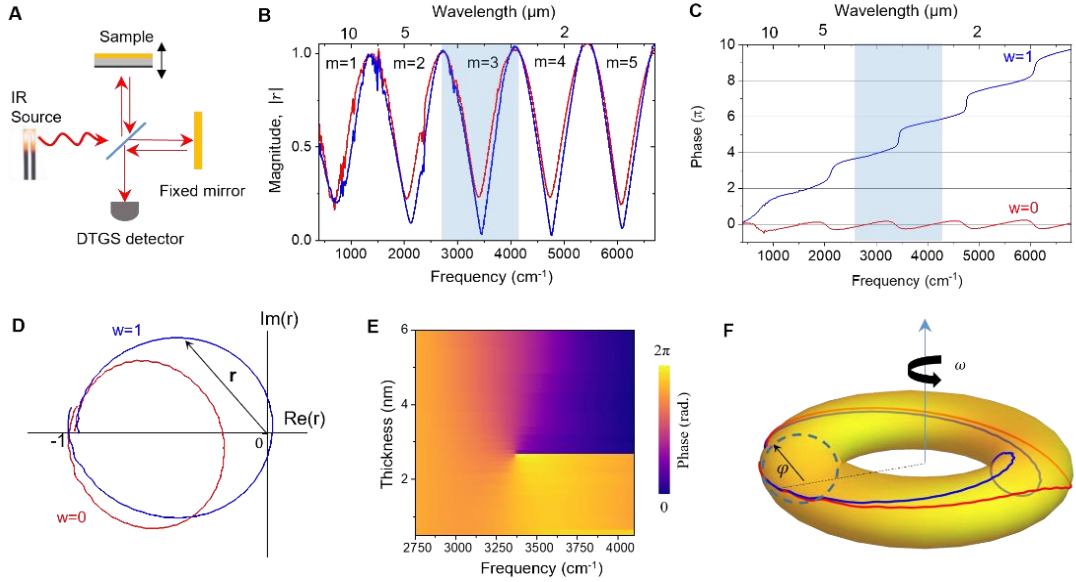

**Fig. 2. Spectroscopic Assessment of Topological Phase Transition: A**, Experimental configuration employing a phase-sensitive Fourier-transform infrared spectrometer, with the sample functioning as a dynamic mirror within the Michelson interferometer. **B**, Magnitude $|r(\omega)|$, and **C**, phase $\varphi(\omega)$ spectra of the Fresnel reflection coefficient for trivial (red curve, w=0, t=2.0 nm) and nontrivial (blue curve, w=1, t=2.7 nm) surfaces, revealing periodic resonance modes possessing a free spectral range of 1325 cm$^{-1}$. **D**, A complex representation of the measured reflection coefficient $r(\omega)$ depicts two distinct winding numbers, 0 and 1, for thicknesses when t<t$_c$ and t>t$_c$, respectively. **E**, Experimentally obtained map of the reflection phase, highlighting a singularity endowed with a topological charge of +1. **F**, Spectrum of the reflection phase mapped on to the surface of a torus here, $\varphi$ is the reflection phase and the rotation over $\phi$ represents frequency over a free spectral range of the cavity. The red and the blue curves are not isotopic on the surface of torus indicating different topologies.

On the most fundamental level, the topological origin of perfect emissivity at critical coupling reveals itself when considering the interface between topologically distinct regions(*21–23*) – such as between a sub-critically and a super-critically coupled cavity. This is because the transition of a topological invariant such as the winding number across such a boundary requires the presence of a phase singularity with $r(\omega) = 0$ at the interface. Due to reciprocity, this line singularity in the absorption is accompanied by a topologically protected interface state with perfect thermal emissivity. We implement such a localised edge state of high-intensity thermal emission on a low emissivity surface by controlling the film thickness $t$ in a stepwise way such that topologically inequivalent areas are formed, whose boundaries will be shown to confine the thermal emission. Furthermore, this line singularity would support chiral waveguide states that can be excited by an external light source(*24*).

To visually capture these boundaries, we fabricated an optical surface on a silicon wafer characterized by two distinct reflection topologies establishing a visible boundary at the wafer's centre (Fig. 3A). To render the interface emission with a far infrared camera, we calibrated the dielectric layer's thickness, aligning the resonance wavelength (~10µm) with the camera's spectral window (8-14 µm). Despite both metal thicknesses yielding low thermal emissivity

(0.2 and 0.3), a pronounced, unpolarized thermal emission was observed at their interface. It's important to mention that one side of the boundary functions as an electrical conductor ($\varphi = \arg[r(\omega)] = \pi$), while the opposite side operates as a magnetic conductor ($\varphi = \arg[r(\omega)] = 0$), each exhibiting low thermal emissivity. However, the interface itself serves as a dipole due to the out-of-phase response on both sides when the topologies differ. This thermal emissivity at the boundary emerges exclusively when there's a change in the winding number across the boundary (see Fig. S4). Additionally, we note that the thermal radiation emitted from these modes is unpolarized (inset in Fig. 3D). Given that these lines of perfect emission ($\varepsilon(\omega_0) = 1$) at the boundary are protected by the reflection topology, they exhibit robustness against local perturbations and defects. This resilience is shown by a more intricate shape — the map of the United Kingdom (defined via photolithography and metallisation) — realized by different metal thicknesses inside and outside of the country's borders. A continuous line of thermal emission, tracing the map's boundary, is clearly observable (Fig. 3D). As can be seen from the enlarged image in Fig. 3D, which shows a small island surrounded by a line of thermal emission at its boundary. The interface state follows the complicated boundary down to the spatial resolution of the infrared camera (~25 µm) and the fidelity of the photolithography process.

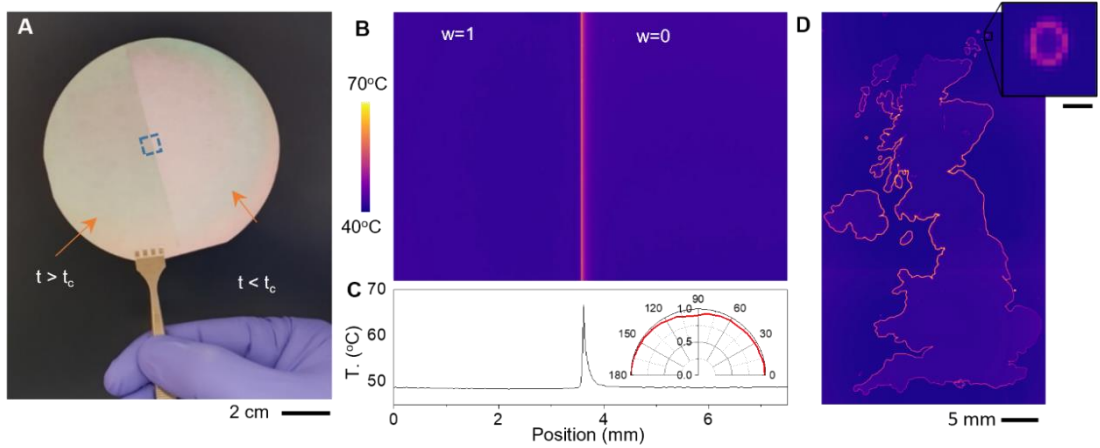

**Fig. 3. Localised thermal emission from a topological interface**: **A**, Photograph of the sample, which includes a boundary between topologically distinct domains. **B**, IR thermograms of the sample, illustrating intense thermal emission localized at the boundary. **C**, Profile of the recorded temperature across the boundary, revealing localised emission. While both domains exhibit low thermal emissivity, the boundary itself presents an emissivity value of 1. The inset details the negligible polarization dependence of the emissivity. **D**, IR thermogram displaying thermal radiation along the boundary of a complex shape (map of the United Kingdom). Here, the different domains on either side of the boundary are realised by trivial and nontrivial topological phases, respectively, with the thermal radiation continuously tracing the map's contour. The inset provides a magnified view of an island, surrounded by a continuous boundary mode. The scalebar in the inset is 100 µm.

To precisely probe the spatial distributions of localized emission, we employed nanoIR photothermal spectroscopy(*25*, *26*), facilitated by an atomic force microscope integrated with a broad-band tuneable infrared laser (Bruker NanoIR3-s). This technique yields direct measurements of nanoscale infrared absorption (see Fig. S5). The AFM cantilever registers the photothermal expansion, an indicator directly proportional to IR absorption, thereby offering insights into local IR emissivity. We first recorded the resonant enhanced contact mode AFM-IR image (Fig. 4B) at the boundary separating regions with top metal layers of 0.5 nm and 30 nm. We observed that the localized absorption at the boundary exhibits an exceptionally narrow spatial distribution (approximately 1 µm, or $\sim\lambda/10$). This distinctive mode was present for both transverse electric (TE) and transverse magnetic (TM) polarized lasers excitations. After these observations, we proceeded with the deposition of an additional 10 nm of metal onto the sample effectively increasing the top metal thickness in the regions to 10.5 nm and 40 nm, respectively, surpassing the critical thickness ($t_c$=2.8 nm) and rendering their reflection topologies identical. Under these modified conditions, the previously detected edge mode disappeared (Fig. 4C), demonstrating that these protected modes manifest themselves exclusively at topological boundaries where the winding number makes a step change.

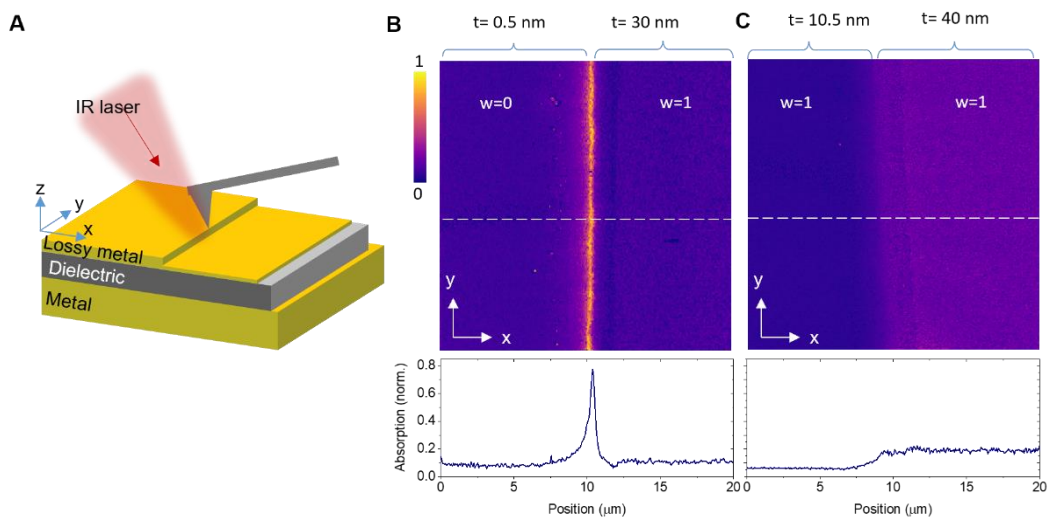

**Fig. 4. NanoIR spectroscopy of the topological edge mode**: **A**, Schematic drawing of the sample geometry and the AFM-IR setup used to measure the local force generated by the photothermal expansion. **B, C** Resonant enhanced contact mode AFM-IR images and IR absorption profiles of a boundary between topologically different (**B**) and identical (**C**) domains. The images were obtained at the excitation wavenumber of 850 cm$^{-1}$.

The hallmark of topological boundaries is their ability to support chiral edge modes. However, with our experimental setup, we cannot excite and visualize these modes. To investigate this feature, we performed 3D finite difference simulations, by placing a chiral dipole emitter positioned in the near field of the boundary. Our simulations shows that the topological boundary between domains with different winding numbers, supports counterpropagating chiral edge modes (Fig. S6-S7 and Supplementary Videos 1-3). These localised modes show strong spin-momentum locking resembling the edge modes found in photonic metasurfaces with band inversion(*27–30*), and to non-Hermitian line modes (*22*). Since the time-reversal symmetry is not broken for our system, it is important to note that these interfaces support two counterpropagating edge modes which could be backscattered with a certain type of defect(*31*) (See Figs. S10-S11). Future work should aim observation of these chiral edge modes not only at infrared but also at extending the wavelength range to near infrared or even visible. Furthermore, the presence of these modes is contingent on the balance between inherent and radiative losses in the cavity. Adjusting the spatial variation of material's loss through an external field could lead to the adaptable paths for these edge modes, thus facilitating the directed transmission of infrared signals. This property could be achieved by materials with tuneable infrared absorption enabling reconfigurable routing.

To conclude, we present a novel topological framework for the design of thermal emitters. Our results show that a state of perfect absorption emerges from a topological phase transition characterized by a change in the winding number i.e. a topological invariant. Despite the origin of thermal radiation being spontaneous emission from uncorrelated sources, the emergence of topological features provides fresh insights and new design strategies for thermal photonics(*32*). Unlike conventional topological photonic systems which relies on topology of the band structure of photonic crystals, our approach does not require any periodic structure to explore topological features. Stepping away from the constraints of periodicity would enable broader applications of topological phenomena.

# ACKNOWLEDGEMENTS

**Funding:** This work was funded through the European Research Council through ERC-Consolidator grant 682723 and ERC PoC grant (Funded by EPSRC EP/X027643/1). SKO and AK acknowledge support from the Air Force Office of Scientific Research (AFOSR) Multidisciplinary University Research Initiative (MURI) Award on Programmable systems with non-Hermitian quantum dynamics (Award FA9550-21-1-0202) and the Air Force Office of Scientific Research (AFOSR) Award FA9550-18-1-0235 and FA9550-22-1-0431. We thank Henry Royce Institute for Advanced Materials for the use of scanning probe microscopy facility.

**Author contributions:** MSE and CK conceived the idea. MSE fabricated the samples. MSE and CK carried out the experiments. KD performed NanoIR measurements. AK and SS performed the electromagnetic simulations. GB performed the transfer matrix calculations. AP and AK provided theoretical support. MSE, SR, SKO and CK analysed the data and wrote the manuscript with input from all authors.

**Competing interests:** Authors declare no competing financial interests.

**Data and code availability:** Data and code are available from the authors upon request.